\documentclass[prd]{revtex4}
\usepackage{color,graphicx}

\usepackage{epsf}
\usepackage{amsbsy}
\usepackage{amssymb}
\usepackage{amsmath}

\def\X{{\mathrm{x}}}
\def\Y{{\mathrm{y}}}

\def\n{{\rm n}}
\def\p{{\rm p}}

\def\bep{\bar{\varepsilon}}
\def\bB{\bar{\mathcal{B}}}
\def\bBp{\bar{\mathcal{B}}'}

\newcommand{\be}{\begin{equation}}
\newcommand{\ee}{\end{equation}}
\newcommand{\bear}{\begin{eqnarray}}
\newcommand{\eear}{\end{eqnarray}}

\begin{document}

\title{On the oscillations of dissipative superfluid neutron stars}

\author{N. Andersson$^1$, K. Glampedakis$^{2,3}$ and B. Haskell$^1$}

\affiliation{$^1$School of Mathematics,
University of Southampton, Southampton SO17 1BJ, United Kingdom}

\affiliation{
$^2$SISSA, via Beirut 2-4, 34014 Trieste, Italy}

\affiliation{$^3$ Theoretical Astrophysics, Auf der Morgenstelle 10, University of Tuebingen, Tuebingen D-72076, Germany}

\begin{abstract}
\begin{center}
We investigate the oscillations of slowly rotating
superfluid stars, taking into account the vortex mediated mutual friction force that
is expected to be the main damping mechanism in mature neutron star cores.
Working to linear order in the rotation of the star, we consider both the
fundamental f-modes and the inertial r-modes. In the case of the (polar) f-modes, 
we work out an analytic approximation of the
mode which allows us to write down a closed expression for the mutual friction damping timescale.
The analytic result is in good agreement with previous numerical results obtained using an energy integral argument.
We  extend previous work by considering the full range of permissible values for the vortex drag, 
e.g. the friction between each individual vortex and the electron fluid. This leads to the first ever results for the f-mode 
in the strong drag regime.  
Our estimates provide useful insight into the dependence on, and relevance of, various equation
of state parameters. In the case of the (axial) r-modes, we confirm the existence of two classes of
modes. However, we demonstrate that only one of these sets remains purely axial in more realistic neutron star models.
Our analysis lays the foundation for companion studies of the mutual friction damping of the
r-modes at second order in the slow-rotation approximation, the first time evolutions for superfluid neutron star perturbations  
and also the first detailed attempt at studying
the dynamics of superfluid neutron stars with both a relative rotation between the components and
mutual friction. 
\end{center}
\end{abstract}


\maketitle

\section{Introduction}

Neutron stars have a complex interior structure. With core densities reaching several times the nuclear saturation density,
these objects require an understanding of physics that cannot be gained from laboratory experiments. This
makes the modelling of neutron star dynamics
an interesting challenge. On the one hand, one has to consider exotic physics that is, at best, poorly constrained.
On the other hand, one may ask to what extent observations can distinguish between
different possible models. An excellent example of this interplay concerns the possibility that the quarks may deconfine in the
high density region. If this is the case, it will have a considerable effect on transport properties associated with
viscosity and heat conductivity. In fact, such a quark core is expected to be a colour superconductor \cite{alford}.
The dynamics of this exotic state of matter, and the relevance of its different possible phases, is not yet
certain. In order to improve our understanding of this problem, we need to build more precise stellar models and
study, for example, their oscillation properties in detail. In this context, considerable attention has been focused
on the inertial r-modes of a rotating star. The r-modes are interesting because they can be driven unstable by the
emission of gravitational radiation, see \cite{nk,nareview} for literature reviews. The r-mode instability window is, however,
 sensitive to the physics of the neutron star interior. Since the bulk and shear viscosities are
quite different in a quark core,  compared to ``normal'' npe matter, one may hope to use observations to constrain the theory, see
\cite{alford} for a discussion of the relevant literature.
In absence of a direct detection of an r-mode gravitational-wave signal, this analysis would have to be
based on the nature of the instability window. The idea would be that, if an observed neutron star spins at a rate
that would place it inside a predicted instability region, one may be able to rule out this particular theoretical model.
Of course, this argument comes with a number of caveats. It could, for example, be that  additional  physics
places a stronger constraint on the r-modes than the considered mechanisms. Inevitably, this becomes a ``work in progress'' where
improved theoretical models are tested against better observational data.

In order to consider ``realistic'' neutron stars, it is important to appreciate the relevance of superfluidity.
A neutron star is expected to contain a number of superfluid/superconducting components \cite{haensel}, and it is
crucial to understand to what extent this affects the stars oscillation properties. It is well established that the
behaviour of a superfluid system can differ significantly from standard hydrodynamics. The most familiar low-temperature system is, perhaps,
He$_4$, which exhibits superfluidity below a critical temperature near 2~K. Experimentally, it has been demonstrated that
this system is very well described by the Navier-Stokes equations above the critical temperature. Below the critical temperature
the behaviour is very different, and a ``two-fluid'' model is generally required (see \cite{helium} for a very recent discussion).
Superfluid neutron stars are, to some extent, similar. The second sound in Helium is analogous to a set of, more or less
distinct, ``superfluid'' oscillation modes \cite{epstein,mendell1,ac01} in a neutron star.
These additional modes arise because the different components of a superfluid system are allowed to move ``through'' each other.
The dissipation channels in a superfluid star are also quite different. Basically, the superfluid flows without friction.
In the outer core of a neutron star, which is dominated by npe matter, one expects the neutrons to be superfluid while
the protons form a superconductor. As a  result, the shear viscosity is dominated by e-e scattering \cite{cutler,npa}.
The bulk viscosity, which is due to the fluid motion driving the system away from chemical equilibrium and the resultant
energy loss due to nuclear reactions, is also expected to be (exponentially) suppressed in a superfluid \cite{haensel}.
These effects have direct implications for the damping of neutron star oscillations, and play a key role in determining the
r-mode instability window for a mature neutron star. This is, however, not the end of the story.  A superfluid exhibits an additional
dissipation mechanism, usually referred to as  ``mutual friction". The mutual friction is due the presence of vortices in a rotating
superfluid. In a neutron star core, the electrons can scatter dissipatively off of the (local) magnetic field of each vortex (see \cite{als,mendell,trev1}
for discussions and references). This effect may dominate the damping of realistic neutron star oscillation modes.

The basic requirements of a rudimentary model for superfluid neutron star oscillations should now be clear. One must account for the
additional dynamical degree(s) of freedom, and also account for the mutual friction damping. This is obviously only a starting point, but the
problem is sufficiently complicated that one may want to proceed with care.
There has already been a number of studies of
dissipative superfluid oscillations. The area was pioneered by Lindblom and Mendell who, in particular, demonstrated that the
gravitational-wave instability of the fundamental f-modes would be suppressed in a superfluid star \cite{lm95}. Following the
discovery of the r-mode instability, they also provided the first accurate estimates of the relevance of the mutual friction
for these modes \cite{lm00}. Similar results were subsequently obtained by Lee and Yoshida \cite{yl1}.  These studies provide
important assessments of the relevance of the mutual friction damping.
There are, however, a number of reasons why we need to return to this problem. Most importantly, we want to consider more realistic
neutron star models, including finite temperature effects, magnetic fields and the possible presence of exotic (hyperon and/or quark)
cores. The additional physics brings additional complications, like additional fluid degrees of freedom, boundary layers at phase-transition interfaces
and fundamental issues concerning dissipative multifluid systems \cite{helium,monster}. We also need to move away from the assumption that the vortex drag,
which leads to the mutual friction, is weak. Strong arguments suggest that this is not going to be the case when the protons form a type~II superconductor
and there are magnetic fluxtubes present in the system \cite{sauls,ruderman,link03}. The neutron vortices may be ``pinned'' to the fluxtubes leading to a strong
drag regime. The strong drag problem has only  been considered recently \cite{precess,kostas}, and the first results demonstrate the presence of a
new instability in systems where the two components rotate at different rates. This instability, which may be relevant for the understanding of pulsar glitches \citep{kostas}, provides
a direct demonstration that the dynamics in the strong drag regime may be both complicated and interesting.
The present investigation lays the foundation for future work in this direction by allowing for strong drag. In particular, we retain the dynamic
contribution to the mutual ``friction'' that has previously been neglected as a matter of course.

\section{The two-fluid equations}

Our discussion is based on the standard two-fluid model for neutron star cores \cite{prix,monster}. That is,
we consider two dynamical degrees of freedom loosely speaking representing the superfluid neutrons
(labeled $\n$) and a charge-neutral conglomerate of  protons and electrons (labeled $\p$).
Assuming that the individual species are conserved, we have the
usual conservation laws for the mass densities $\rho_\X$,
\be
\partial_t \rho_\X + \nabla_i (\rho_\X v_\X^i) = 0
\ee
where the constituent index x may be either p or n. Meanwhile, the equations of momentum balance can be written
\be
(\partial_t + v_\X^j \nabla_j ) (v^\X_i+\varepsilon_\X w^{\Y\X}_i) +\nabla_i (\tilde{\mu}_\X+\Phi)
+ \varepsilon_\X w^j_{\Y\X} \nabla_i v^\X_j= f^\X_i/\rho_\X
\label{Eulers}\ee
where the velocities are $v_\X^i$, the relative velocity is defined as $w_{\X\Y}^i = v_\X^i-v_\Y^i$ and
 $\tilde{\mu}_\X=\mu_\X/m_\X$ represents the 
chemical potential (we will assume that $m_\p=m_\n$ throughout this paper). $\Phi$ represents the gravitational potential, and the parameter
$\varepsilon_\X$ encodes the non-dissipative entrainment coupling between the fluids \cite{prix,monster}.
The force on the right-hand side
of (\ref{Eulers}) can be used to represent various other interactions, including dissipative terms \cite{monster}.

In the following we will focus on the vortex-mediated mutual friction. Assuming that the two fluids  exhibit solid body rotation
we have a force of form \cite{trev1} (see also \cite{als,mendell})
\be
f^\X_i = 2  \rho_\n \mathcal{B}' \epsilon_{ijk}\Omega^j w_{\X\Y}^k
+ 2  \rho_\n  \mathcal{B} \epsilon_{ijk}\hat{\Omega}^j \epsilon^{klm} \Omega_l w_m^{\X\Y}
\label{mf}\ee
Here, $\Omega^j$ is the angular frequency of the neutron fluid (a hat represents a unit vector).
The mutual friction parameters are intimately related to the induced friction on the vortex. The latter is
often described in terms of a dimensionless ``drag'' parameter $\mathcal{R}$ such that
\be
\mathcal{B}' = \mathcal{R} \mathcal{B} = { \mathcal{R}^2 \over 1 + \mathcal{R}^2 }
\label{bvsr}\ee
In the standard picture, the mutual friction is due to the scattering of electrons off of the array
of neutron vortices. This leads to $\mathcal{R} \ll 1$, i.e., $\mathcal{B}'\ll\mathcal{B}$, and hence the first term in the mutual
friction force can be ignored. There are, however, good arguments for why the problem may be in the opposite
regime. In particular if one considers the interaction between the fluxtubes in a type~II proton superconductor
and the neutron vortices \cite{sauls,ruderman,link03}. Then one would expect to be in the strong drag regime
where $\mathcal{R}\gg1$, i.e., $\mathcal{B}'\approx 1$ while $\mathcal{B}$ remains small.  Superfluid oscillations in
this regime have not previously (with the
exception of \cite{kostas}) been considered.

Anyway, from (\ref{mf}) we see that the mutual friction will not be present in a non-rotating star. This is obvious
since there would then be no vortices in the first place. Of course, any non-trivial motion of the
superfluid neutrons  leads to vortex generation. This means that a generic perturbation of a non-rotating star will be associated with
a local vorticity which could lead to mutual friction. However, in this context the resulting
mutual friction interaction would require a perturbative calculation to be carried out to second order. As far as we are aware,  such calculations have not yet been attempted.
It may be an interesting problem for the future.

\section{The perturbation equations}

\subsection{Decoupling the degrees of freedom}

If we want to consider the effects of mutual friction
we  need to consider rotating stars. To keep the problem tractable (at least initially) we
assume that the background configuration is such that the two fluids rotate together. Perturbing the equations of motion and working in a frame rotating with $\Omega^j$ we then have
\be
\partial_t (\delta v^\X_i + \varepsilon_\X\delta w_i^{\Y\X}) + \nabla_i (\delta \tilde{\mu}_\X + \delta \Phi)
+ 2 \epsilon_{ijk}\Omega^j \delta v_\X^k = \delta ( f^\X_i/\rho_\X)
\label{perteul}\ee
and
\be
\partial_t \delta \rho_\X + \nabla_j (\rho_\X \delta v_\X^j) = 0
\ee
where $\delta$ represents an Eulerian variation.

From previous work on superfluid neutron star perturbations (and indeed the large body of work on superfluid Helium)
we know that the problem has two "natural" degrees of freedom, see for example \cite{epstein,mendell1,ac01,prix1,prix2,trev2}. One of the degrees of freedom
represents the total mass flux.
Introducing
\be
\rho \delta v^j = \rho_\n \delta v_\n^j + \rho_\p \delta v_\p^j
\ee
and combining the two Euler equations we find that
\be
\partial_t \delta v_i +  \nabla_i \delta \Phi + { 1 \over \rho} \nabla_i\delta p - { 1 \over \rho^2} \delta \rho \nabla_i p
+ 2 \epsilon_{ijk}\Omega^j \delta v^k = 0
\label{Eulav}\ee
where  $\rho = \rho_\n+\rho_\p$ and the pressure is obtained from~\footnote{Caution: In the general case when the background fluids are not co-rotating, there
will be additional  contributions associated with the entrainment in this relation.}
\be
\nabla_i p = \rho_\n \nabla_i \tilde{\mu}_\n + \rho_\p \nabla_i \tilde{\mu}_\p
\ee
In deriving this relation we have used
\be
\rho_\n \nabla_i \delta\tilde{\mu}_\n + \rho_\p \nabla_i \delta \tilde{\mu}_\p = \nabla_i \delta p - \delta \rho \nabla_i \tilde{\mu}
=  \nabla_i \delta p - { 1 \over \rho} \delta \rho \nabla_i p
\ee
where it has been assumed that the two fluids are in chemical equilibrium in the background. That is, we have
$\tilde{\mu}_\n  =\tilde{\mu}_\p=\tilde{\mu}$.
We also have the usual continuity equation
\be
\partial_t \delta \rho + \nabla_j (\rho \delta v^j) = 0
\label {conav}\ee

At this point we have two equations which are identical to the perturbation equations for a single fluid
system. It is particularly notable that (\ref{Eulav}) does not have a force term. This follows
immediately from the fact that we are only considering the mutual friction interaction. In other situations, say
including shear viscosity, we would no longer have a homogeneous equation.

Of course, we are considering a two-fluid problem and there is a second degree of freedom to take into account.
To describe this, it is natural to consider the difference in velocity. Thus, we introduce
\be
\delta w^j = \delta v_\p^j - \delta v_\n^j
\ee
Combining the two Euler equations in the relevant way we have
\bear
(1-\bep) \partial_t \delta w_i + \nabla_i \delta \beta +
2 \bBp \epsilon_{ijk}\Omega^j \delta w^k - \bB \epsilon_{ijk}\hat{\Omega}^j \epsilon^{klm} \Omega_l \delta w_m = 0
\label{Euldi}\eear
Here we have defined
\be
\delta \beta = \delta \tilde{\mu}_\p - \delta \tilde{\mu}_\n
\ee
which represents the (local) deviation from chemical equilibrium induced by the perturbations. We have also introduced
the simplifying notation
\be
\bep = \varepsilon_\n/x_\p \ , \qquad \bBp = 1-\mathcal{B}'/x_\p \ , \qquad
\bB = \mathcal{B}/x_\p
\ee
where $x_\p=\rho_\p/\rho$ is the proton fraction.
Again,  equation \eqref{Euldi} does not couple the different degrees of freedom.

The coupling is entirely due to the second continuity equation. It is 
natural to use the proton fraction to complement the total density $\rho$. Then we find that
\be
\partial_t \delta x_\p + { 1 \over \rho} \nabla_j \left[ x_\p (1-x_\p) \rho \delta w^j \right] +  \delta v^j \nabla_j x_\p = 0
\label{condi}\ee
This equation shows that the two dynamical degrees of freedom are explicitly coupled
\underline{unless} the proton fraction is constant. This fact has already been
pointed out by Prix and Rieutord \cite{prix1}.

Before moving on, it is worth asking to what extent it is possible to find solutions that are
\underline{purely} co-moving, i.e. for which $\delta w^j= \delta \beta=0$. From the above equations it is easy to see that 
such a solution would have to satisfy
\be
\partial_t \delta x_\p  +  \delta v^j \nabla_j x_\p = 0
\ee
This condition is trivially satisfied if the proton fraction is uniform. In addition, it will be satisfied for fluid motion 
that has (for a spherical background configuration) no radial component \underline{and} also do not lead to variations in $\delta x_\p$. 
Are there oscillation modes with this character? Indeed, to leading order in the slow-rotation approximation the standard r-mode 
satisfies these criteria. It is purely axial and the associated density perturbations appear at order $\Omega^2$. However, in general 
we do not expect to find any oscillations of a ``realistic'' neutron star model to be purely co-moving. This means that a generic
neutron star oscillation mode will be affected by mutual friction. 

\subsection{Boundary conditions}

To completely specify the perturbation problem, we need boundary conditions. At the centre of the star
we simply require that all variables are regular. The surface of the star is somewhat more complex.
In reality one
does not expect the superfluid region to extend all the way to the surface. A real neutron star will
always be covered by a single fluid envelope (eg. the outer parts of the elastic crust). However, for simplicity
we do not want to deal with the various interfaces in the present analysis \cite{lm00,layers,lin}.
Instead we will consider stars with a two-fluid surface, which is  obviously somewhat artificial.

A reasonable approach is to assume that the perturbed star has a unique surface. That is,
let the two perturbed fluids move together (in the radial direction) at the surface.
In the two-fluid problem we have two distinct Lagrangian displacements $\xi^j_\X$ \cite{grosart}.
These follow from
\be
\partial_t \xi_\X^i =  \delta v_\X^i + \xi_\X^j \nabla_j v_\X^i + v_\X^j \nabla_j \xi_\X^i
\ee
We have assumed that the two fluids corotate in the background, i.e. we have $v_\n^i = v_\p^i$.
If we also impose that there is a common surface, then we have $\xi_\n^r=\xi_\p^r$ at $r=R$
and it follows that we should require
\be
\delta w^r= \delta v_\p^r  -\delta v_\n^r = 0 \ , \qquad \mbox{ at } r=R
\ee
From \eqref{Euldi} we see that this implies that, for a non-rotating configuration we must also have
\be
\partial_r \delta \beta = 0  \ , \qquad \mbox{ at } r=R
\ee
When we determine the rotational corrections to the f-mode we will impose this condition also at first
slow-rotation order. This is not entirely consistent, cf. \eqref{Euldi}, but it is straightforward to relax this condition 
should it be required.

If the two fluids move together at the surface, it also follows that
\be
\delta p + \rho \xi^j \nabla_j \tilde{\mu} = \delta p + \xi^r \partial_r p = \Delta p = 0 \ , \qquad \mbox{ at } r=R
\ee
where $\xi^r = \xi_\n^r =\xi_\p^r$ at the surface. This is the usual single fluid
condition of a vanishing Lagrangian pressure variation $\Delta p$.

\subsection{A bit of chemistry}

Consider the various equations that we have written down. At this point the two
degrees of freedom $[\delta v_i, \delta p]$ and $[\delta w_i, \delta \beta]$ only couple explicitly through
(\ref{condi}). In fact,  if we assume that the two fluids are incompressible,
then there is no coupling at all. Since the mutual friction only enters the problem via (\ref{Euldi}),
it is thus the case that any incompressible dynamics in the $[\delta v_i, \delta p]$ sector will be unaffected
by mutual friction. This shows that, if we are interested in the effect of mutual friction on (say) the f-mode
oscillations of a star it is not meaningful to consider an incompressible model. We know already from the outset that we
would only find the mutual friction effects on the counter-moving ``superfluid" modes. This may  be an interesting
problem, but it is not our main motivation here.

For compressible models, the two degrees of freedom also couple indirectly. Basically, we need to use the equation
of state to relate $[\delta p,\delta \beta]$  to $[\delta \rho, \delta x_\p]$. For models where the two
fluids co-rotate in the background we can use~\footnote{Note that there will be entrainment contributions here in the general case when the
two background fluids are not co-rotating.}
\be
\delta \rho = \left( { \partial \rho \over \partial p} \right)_\beta \delta p +
\left( { \partial \rho \over \partial \beta} \right)_p \delta \beta
\ee
and
\be
\delta x_\p = \left( { \partial x_\p \over \partial p} \right)_\beta \delta p +
\left( { \partial x_\p \over \partial \beta} \right)_p \delta \beta
\ee
Using these relations, or their ``inverse'', we see that the two degrees of freedom couple in a more subtle way.
If we choose to reduce the problem by  eliminating $\delta p$ and $\delta \beta$ then  the coupling arises
through the boundary conditions and the Euler equations.
If, on the other hand, we eliminate $\delta \rho$ and $\delta x_\p$ then
the coupling enters through the continuity equations.

\section{Dissipation integrals}

In order to estimate the damping associated with various dissipation mechanisms one can either (i) account for the
dissipative terms in the equations of motion and solve for the damped modes directly, or (ii) solve the
non-dissipative problem and use an energy integral argument to estimate the damping rate. In the typical situation when the damping
is very slow the second strategy should be reliable. Indeed, all studies of damped neutron star oscillations have
used this approach (see \cite{nareview} for a discussion). Given this, it is natural to pause and consider the energy integral
approach to the mutual friction problem. 


\subsection{The conserved energy}

To work out a suitable energy associated with a given perturbation, we first multiply (\ref{perteul}) with
$\rho_\X \delta \bar{v}_\X^i$ (where the bar represents complex conjugation). Then we add the result to its
complex conjugate.  Combining the individual contributions from the neutron and proton fluids
and integrating over the star we
find that, when $f_i^\X=0$, the result is a total time derivative of two terms. The first term is the
 ``kinetic energy'', which follows from
\be
E_k = { 1 \over 2} \int\left[ (\rho_\n-2\alpha) |\delta v_\n|^2 + 4\alpha \mbox{Re} ( \delta \bar{v}_\n^i \delta v^\p_i)
+ (\rho_\p-2\alpha) |\delta v_\p|^2 \right] dV
\label{kinetic}\ee
where $2\alpha = \rho_\X \varepsilon_\X$.
Alternatively, expressing this in the co- and countermoving variables, we have 
\be
E_k = \frac{1}{2} \int  \rho \left[ |\delta v|^2 + (1-\bar{\varepsilon}) x_\p (1-x_\p) |\delta w|^2  \right] dV
\ee

The ``potential'' energy requires a bit more work. Using the divergence theorem and the continuity equation one can show that we need
\be
\int \rho_\X [ \delta \bar{v}_\X^i \nabla_i (\delta \tilde{\mu}_\X + \delta\Phi) + \mbox{c.c}] dV = 
\int  \rho_\X [ \delta \bar{v}_\X^i (\delta \tilde{\mu}_\X + \delta\Phi) + \mbox{c.c}] \hat{e}^r_i  dS +
\int [ (\delta \tilde{\mu}_\X + \delta\Phi) \partial_t \delta \bar{\rho}_\X + \mbox{c.c}] dV
\ee
(c.c. represents the complex conjugate). The surface term vanishes if $\rho_\X\to 0$ at the surface of the star.
It also vanishes for modes that have no radial component. Adding the contributions for the neutron and proton fluids we see that we need
\be
\int \delta\Phi \partial_t \delta \bar{\rho} dV =
{ 1 \over 4 \pi G} \int  \delta\Phi \partial_t \nabla^2 \delta \bar{\Phi} dV 
={ 1 \over 4 \pi G} \int  \delta\Phi \hat{e}_r^i \nabla_i \partial_t \delta \bar{\Phi} dS - { 1 \over 4 \pi G} \int  (\nabla_i \delta\Phi)\partial_t (\nabla^i \ \partial_t \delta \bar{\Phi}) dV
\ee
Again, one can argue that the surface term vanishes.

Finally, we have terms of form
\begin{displaymath}
\int \delta \tilde{\mu}_\X \partial_t \delta \bar{\rho}_\X dV
\end{displaymath}
It is natural to express these in terms of the perturbed densities
using (this is valid for co-rotating background models only)
\be
\delta \tilde{\mu}_\X = \left({\partial \tilde{\mu}_\X \over \partial \rho_\X} \right)_{\rho_\Y} \delta \rho_\X + \left({\partial \tilde{\mu}_\X \over \partial \rho_\Y} \right)_{\rho_\X} \delta \rho_\Y
\ee

Adding all the terms together we find that the ``potential energy'' follows from
\be
E_p = { 1 \over 2} \int \Big[ \left({\partial \tilde{\mu}_\n \over \partial \rho_\n} \right)_{\rho_\p} | \delta \rho_\n |^2 + 2
\left({\partial \tilde{\mu}_\n \over \partial \rho_\p} \right)_{\rho_\n}
\mbox{ Re } (\delta \rho_\p \delta \bar{\rho}_\n)
+  \left({\partial \tilde{\mu}_\p \over \partial \rho_\p} \right)_{\rho_\n} | \delta \rho_\p |^2
-  { 1 \over 4 \pi G} |  (\nabla_i \delta\Phi)|^2 \Big] dV
\ee
or
\be
E_p = \frac{1}{2} \int  \left\{ \rho \left ( \frac{\partial \rho}{\partial p} \right )_\beta |\delta h |^2 
+ \left (  \frac{\partial \rho}{ \partial \beta} \right )_p \left[ 2 \mbox{Re} ( \delta h \delta \bar{\beta} ) 
+  |\delta \beta |^2 \right]  -\frac{1}{4\pi G}  | \nabla_i \delta \Phi |^2  \right\} dV
\ee

With these definitions it follows that the  total ``energy'' is conserved, i.e.
\be
\partial_t E = \partial_t (E_k + E_p) = 0
\ee
when $f_i^\X=0$. These energy expressions are equivalent to those used by Lindblom and Mendell \cite{lm95}.

\subsection{Mutual friction}

Even though we will include the mutual friction terms in the equations
of motion, it is useful to work out the corresponding dissipation integrals.
After all, this is the way that the mutual friction damping has
traditionally been evaluated \cite{lm95,lm00,yl1} and we want to be able to compare the
two approaches.

First consider the $\mathcal{B}'$ terms. It is easy to show that
these terms are not dissipative.
We find that
\be
2 \partial_t E_{\mathcal{B}'} = 2 \int \mathcal{B}' \epsilon_{ijk}\Omega^j  \Big[ \delta \bar{v}_\n^i \delta w_{\n\p}^k +  \delta \bar{v}_\p^i \delta w_{\p\n}^k + \mbox{c.c.} \Big] dV  = 0
\ee
by symmetry. This result is not surprising. In fact, we see from (\ref{Euldi}) that the $\mathcal{B}'$ terms enter the equations of motion in the same way as the Coriolis force. 
Since the Coriolis terms vanish identically when we multiply each Euler equation with $\delta v_\X^i$ this should be true also for the non-dissipative part of the mutual friction.

Finally, it is  straightforward to show that the dissipative terms lead to
\be
 \partial_t E_{\mathcal{B}} =  \int \rho_\n \mathcal{B}
[ \delta_i^l\delta_j^m - \delta_i^m \delta_j^l]\hat{\Omega}^j \Omega_l [
 \delta \bar{v}_\n^i \delta w^{\n\p}_m +  \delta \bar{v}_\p^i \delta w^{\p\n}_m + \mbox{c.c.} ] dV  = 
  - 2 \int \rho_\n \mathcal{B} \Omega [ \delta_i^m - \hat{\Omega}^m \hat{\Omega}_i]\delta \bar{w}^i_{\p\n}\delta w_m^{\p\n} dV
\label{dE}\ee

Let us now ask how we can use these results to estimate the mutual friction damping timescale. Let us assume that we have a mode solution to the full dissipative problem.
That is, we have a solution with time dependence $e^{i\omega t}$ where $\omega = \omega_r + i/\tau$ such that $\tau$ is the damping timescale. From the fact that the energy is 
quadratic in the perturbations it follows that \cite{il91}
\be
\tau = \left| { 2 E \over \partial_t E} \right|
\ee
Moreover, since the solution satisfies the dissipative equations of motion we also know that
\be
\partial_t E = \partial_t E_{\mathcal{B}}
\ee
Hence, we can equally well use
\be
\tau = \left| { 2 E \over \partial_t E_\mathcal{B}} \right|
\label{tauest}\ee
As long as we are using the complete solution to evaluate this expression, it is an identity. However, in many cases we do not
have access to the solution to the dissipative problem. (If we did, we would not need the energy integrals in the first place.)
In these cases we can still estimate the damping timescale by evaluating the right-hand side of (\ref{tauest}) using the non-dissipative mode solution.
When the damping is sufficiently slow, in the sense that the dissipative terms have a small effect on the eigenfunctions, this estimate should be 
reliable. Of course, one should not expect it to yield exactly the same result as the solution to the full dissipative problem.

\subsection{Gravitational-wave emission}

Finally, let us work out the multipole formulas for
gravitational-wave emission from a two-fluid star.
This exercise is particularly relevant if we are interested in oscillations
that may be driven unstable by gravitational-wave emission \cite{nareview}.
The main motivation for including it here is that it demonstrates
the intuitive result that gravitational waves are only
generated by the co-moving degree of freedom.

 Following \cite{thorne} we need
the mass multipoles
\be
\delta D_{lm} = \int \tau_{00} \bar{Y}_{lm} r^l dV
\approx \int \delta T_{00} \bar{Y}_{lm} r^l dV
\ee
and the current multipoles
\be
\delta J_{lm} = \int (-\tau_{0j}) \bar{Y}_{j,lm}^B dV
\approx \int (-\delta T_{0j}) \bar{Y}_{j,lm}^B dV
\ee
In these expressions $Y_{lm}$ are the standard spherical harmonics and ${Y}_{j,lm}^B$ are the magnetic multipoles \cite{thorne}.

To work these out we start with the usual expression for the two-fluid stress-energy
tensor in general relativity \cite{lin}
\be
T_{\mu\nu} = \Psi g_{\mu\nu} + n_\mu^\n \mu_\nu^\n + n_\mu^\p \mu_\nu^\p
\ee
In the relativistic formulation, see \cite{livrev} for a review and a survey of the literature, the central variables are the fluxes
$n_\X^\mu$. The associated momenta follows from
\be
\mu_\nu^\X = \mathcal{B}^\X n_\nu^\X + \mathcal{A}^{\X\Y} n_\nu^\Y
\ee
Hence, the $\mathcal{A}^{\X\Y}$ coefficients encode the entrainment effect.
Let us now work in the frame of an observer moving with four-velocity
$u^\mu$ such that
\be
u_\X^\nu = [ \gamma_\X , \gamma_\X v_\X^i] \ , \qquad \mbox{with } \qquad
\gamma_\X = ( 1- v_\X^2)^{-1/2}
\ee
where $v_\X^i$ is the associated three-velocity.
Then it follows that
\be
T_{00} = \Psi g_{00} + n_\n^2 \mathcal{B}^\n \gamma_\n^2 + 2 n_\n n_\p
\mathcal{A}^{\n\p} \gamma_\n \gamma_\p + n_\p^2
\mathcal{B}^\p \gamma_\p^2
\ee
We want the Newtonian (low-velocity) limit of this expression. Thus we let
\be
g_{00} \to - 1 \ , \qquad \mbox{ and } \qquad \gamma_\X \to 1
\ee
and we get
\be
T_{00} \approx - \Psi + n_\n^2 \mathcal{B}^\n  + 2 n_\n n_\p
\mathcal{A}^{\n\p} + n_\p^2
\mathcal{B}^\p
\ee
Finally use the definition
\be
\Psi = \Lambda - n_\nu^\n \mu_\n^\nu - n_\nu^\p\mu_\p^\nu \approx
\Lambda +  n_\n^2 \mathcal{B}^\n  + 2 n_\n n_\p
\mathcal{A}^{\n\p} + n_\p^2
\mathcal{B}^\p
\ee
and $\Lambda = - \rho$ to arrive at
\be
T_{00} = \rho \longrightarrow \tau_{00} \approx \delta \rho
\ee
as one would have expected.

For the current multipoles we need
\be
T_{0j} = n_\n^2 \mathcal{B}^\n \gamma_\n^2 v^\n_j + n_\n n_\p
\mathcal{A}^{\n\p}
\gamma_\n \gamma_\p( v^\p_j + v^\n_j ) + n_\p^2 \mathcal{B}^\p \gamma_\p^2 v^\p_j
\ee
In the low-velocity limit, this leads to
\be
T_{0j} \approx n_\n ( n_\n \mathcal{B}^\n  + n_\p
\mathcal{A}^{\n\p} ) v^\n_j  + n_\p (n_\p  \mathcal{B}^\p  v^\p_j+ n_\n
\mathcal{A}^{\n\p} ) v^\p_j
\ee
Finally, we need to write this expression in terms of the Newtonian
variables. This can be done by comparing the momenta,
\be
{ \mu_\n^j \over m} \approx { \mathcal{B}^\n n_\n \over m} v_\n^j
+ { \mathcal{A}^{\n\p} n_\p \over m} v_\p^j = ( 1 - \varepsilon_\n) v_\n^j +
\varepsilon_\n v_\p^j
\ee
 This suggests that we should identify
\be
{ \mathcal{B}^\n n_\n \over m} =  1 - \varepsilon_\n \ , \qquad \mbox{ and } \qquad
{ \mathcal{A}^{\n\p} n_\p \over m} = \varepsilon_\n
\ee
Using analogous expressions for the protons we see that
\be
T_{0j} \approx \rho_\n v^\n_j + \rho_\p v^\p_j
\ee
which leads to (for a co-rotating background)
\be
\tau_{0j} \approx \rho_\n \delta v^\n_j + \rho_\p \delta v^\p_j + \delta \rho v_j  = \rho \delta v_j  + \delta \rho v_j
\ee

These results show that it is only the co-moving degree of freedom
that radiates gravitationally.

\section{Slow rotation perturbation equations}

Let us now return to the problem of oscillating superfluid neutron stars. We will first
 derive the general perturbation equations for a slowly rotating superfluid star.
To do this we expand all variables in spherical harmonics. Since we expect rotation to couple the various multipoles,
we represent the velocity perturbations by the general expressions
\be
\delta v^j = \sum_l \left[ { 1 \over r} W_l Y_l^m \hat{e}^j_r + \left({ 1 \over r^2}  V_l \partial_\theta Y_l^m + { m \over r^2 \sin\theta} U_l Y_l^m
 \right)\hat{e}_\theta^j + { i \over r^2 \sin \theta}  \left( m V_l Y_l^m + U_l \partial_\theta Y_l^m  \right) \hat{e}_\varphi^j  \right]
\ee
and
\be
\delta w^j = \sum_l \left[ { 1 \over r} w_l Y_l^m \hat{e}^j_r + \left({ 1 \over r^2}  v_l \partial_\theta Y_l^m + { m \over r^2 \sin\theta} u_l Y_l^m
 \right)\hat{e}_\theta^j + { i \over r^2 \sin \theta}  \left( m v_l Y_l^m + u_l \partial_\theta Y_l^m  \right) \hat{e}_\varphi^j  \right]
\ee
Note that we represent the ``co-moving'' degree of freedom by the uppercase amplitudes $[W_l,V_l,U_l]$ while the ``counter-moving''
degree of freedom corresponds to the lowercase quantities $[w_l,v_l,u_l]$.
All scalar perturbations are expanded in spherical harmonics, i.e. we have $\delta p = \sum_l \delta p_l Y_l^m$ etcetera.
From now on the sum over $l$ will be implied.

One can use a number of different strategies in writing down the perturbation equations. To some extent this is a matter
of taste. However, in the slow-rotation problem it can be advantageous to work with a set of equations where the
coupling between different multipoles is minimal. The set of equations that we use was chosen using this criterion.
We also decided to use the velocity perturbations as our main variables. This approach is analogous to that used by Lockitch
and Friedman \cite{lockitch} in their analysis of inertial modes of single fluid stars. It is notably different from the two-potential formalism pioneered
by Ipser and Lindblom \cite{ipser}, which was extended to superfluid stars by Lindblom and Mendell \cite{lm00}.

We replace each of the perturbed Euler equations with three equations. The first is the radial component of the vorticity equation that
follows if we take the curl of (\ref{Eulav}) or (\ref{Euldi}). Assuming that the perturbations have a harmonic dependence on time, $\exp(i\omega t)$, we get
\be
[l(l+1)\omega-2m\Omega] U_l Y_l^m + 2\Omega (l+2)(W_l-l V_l)Q_{l+1}Y_{l+1}^m - 2\Omega (l-1) [W_l+(l+1)V_l] Q_l Y_{l-1}^m = 0
\label{vrav}\ee
and
\bear
\left\{ l(l+1)\omega( 1 - \bep) - 2m\Omega \bBp - 2i\Omega[l(l+1)-m^2]\bB \right\} u_l Y_l^m \nonumber \\
+ 2\Omega [ (l+2)\bBp-im\bB ] \left\{  w_l -l v_l \right\} Q_{l+1} Y_{l+1}^m  \nonumber \\
-2\Omega [ (l-1)\bBp+im\bB] \left\{ (l+1) v_l +  w_l  \right\} Q_l Y_{l-1}^m = 0
\label{vrdi}
\eear
In deriving these equation we have made use of the standard recurrence relations
\be
\cos \theta Y_l^m = Q_{l+1} Y_{l+1}^m + Q_l Y_{l-1}^m
\ee
and
\be
\sin \theta \partial_\theta Y_l^m = l Q_{l+1} Y_{l+1}^m - (l+1) Q_l Y_{l-1}^m
\ee
where
\be
Q_l = \left[ {(l-m)(l+m) \over (2l-1) (2l+1)}\right]^{1/2}
\ee
For future reference, note that $Q_m=0$ and $Q_{m+1}^2=1/(2m+3)$.

Next we could use also the $\theta$ (or $\varphi$) components of the
vorticity equation. However, as discussed in \cite{ekman}
there is a slightly simpler alternative. We first of all use a pair of equations analogous to the ``divergence'' equation  in \cite{ekman}. 
These can be written
\bear
\left\{ [ l(l+1)\omega - 2m\Omega] V_l - 2m\Omega W_l -  i l(l+1) [(1-x_\p)\delta \tilde{\mu}^\n_l + x_\p \delta \tilde{\mu}^\p_l]  \right\} Y_l^m \nonumber \\
- 2\Omega l(l+2)U_l Q_{l+1} Y_{l+1}^m - 2\Omega (l^2-1) U_l Q_l Y_{l-1}^m = 0
\label{divav}\eear
and
\bear
&& - \left\{ i l(l+1) \delta \beta_l + 2m\Omega\bBp w_l-[ l(l+1)\omega(1-\bep) -2m\Omega\bBp]v_l \right. \nonumber \\
&& +  2i\Omega \bB \left[ 1 - (l+3) Q_{l+1}^2 +(l-2)Q_l^2 \right] w_l \nonumber \\
&&
\left. + 2i\Omega \bB \left[ m^2 + l(l+3) Q_{l+1}^2 +(l^2-l-2)Q_l^2 \right] v_l
\right\} Y_l^m \nonumber \\
&& - 2\Omega (l+2)(l\bBp+im\bB)u_l Q_{l+1}Y_{l+1}^m
-2\Omega (l-1)[(l+1)\bBp-im\bB] u_l Q_lY_{l-1}^m \nonumber \\
&& -2i\Omega\bB (l+3)(lv_l-w_l)Q_{l+1}Q_{l+2}Y_{l+2}^m
-2i\Omega\bB (l-2)[w_l+(l+1)v_l]Q_l Q_{l-1}Y_{l-2}^m = 0
\label{divdi}\eear

Meanwhile the radial components of the Euler equations lead to
\bear
&& \left\{ i[(1-x_\p) r \partial_r \delta  \tilde{\mu}^\n_l + x_\p r \partial_r \delta \tilde{\mu}^\p_l] + 2m\Omega V_l
-\omega W_l \right\} Y_l^m \nonumber  \\
&+& 2\Omega l U_l Q_{l+1} Y_{l+1}^m - 2 \Omega(l+1) U_l Q_l Y_{l-1}^m = 0
\label{eulrav}\eear
and
\bear
&&
\left\{
ir \partial_r \delta \beta_l  - [ \omega(1-\bep) - 2i \Omega(1 - Q_l^2 - Q_{l+1}^2)\bB w_l
+ 2\Omega [ m\bBp - i ((l+1)Q_l^2-l Q_{l+1}^2) \bB] v_l
\right\} Y_l^m \nonumber \\
&&
+ 2\Omega ( l \bBp+im\bB) u_l Q_{l+1} Y_{l+1}^m - 2\Omega[ (l+1) \bBp-im\bB] u_l Q_lY_{l-1}^m
\nonumber \\
&&
+ 2i\Omega (l v_l -w_l) \bB Q_{l+1} Q_{l+2} Y_{l+2}^m
- 2i \Omega [(l+1)v_l +w_l]\bB Q_l Q_{l-1} Y_{l-2}^m = 0
\label{eulrdi}\eear

Finally, the continuity equations become
\be
i \omega r^2 \delta \rho_l + \partial_r (r \rho W_l) - l(l+1) \rho V_l = 0
\label{cav} \ee
and
\be
i \omega \rho r^2 \delta x_{l} + \partial_r [ x_\p(1-x_\p)r \rho w_l ] - x_\p(1-x_\p) l(l+1) \rho v_l + \rho W_l r \partial_r x_\p = 0
\label{cdi}\ee
This completes the description of the general first order slow-rotation problem.

In order to deduce the relevant recurrence relations from the above equations
we need to recall that we have been implying summation over $l$. That is,
we are  considering relations of form
\be
\sum_l \left[ a_l Q_l Q_{l-1}Y_{l-2}^m + b_l Q_lY_{l-1}^m
+ c_l Y_l^m + d_l Q_{l+1}Y_{l+1}^m + e_l Q_{l+1}Q_{l+2}Y_{l+2}^m \right] = 0
\ee
Using orthogonality of the spherical harmonics, i.e. multiplying by
$\bar{Y}_n^m$ and integrating over the sphere, we obtain the recurrence
relation
\be
a_{n+2} Q_{n+1} Q_{n+2} + b_{n+1} Q_{n+1} + c_n + d_{n-1} Q_{n-1}
+e_{n-2} Q_{n-1} Q_n =0
\ee
Given this result, it is  straightforward to write down recurrence relations
for the various classes of oscillation modes of a rotating superfluid star.
However, since the level of rotational coupling is different for different kinds of modes,
it is not particularly useful to write down the general relations.
Instead, we  focus on two specific examples.

\section{The f-modes}

Let us begin by considering modes that are non-trivial already in a non-rotating star.
Then we first need to solve the non-rotating (and non-dissipative since the
mutual friction damping requires rotation) problem. Simply setting
$\Omega=0$ in our perturbation equations we see that the polar and axial degrees of
freedom decouple (as they should). It is also clear, cf.
(\ref{vrav}) and (\ref{vrdi}), that there will not exist
any purely axial modes in the non-rotating case.
This means that we can make the Ansatz
\be
\omega = \omega_0 + \omega_1 \Omega
\ee
together with
\be
W_l = W^0_l+ \Omega W^1_l \ , \quad V_l = V^0_l+ \Omega V^1_l \ , \quad U_l =
\Omega U^1_l
\ee
and
\be
w_l = w^0_l+ \Omega w^1_l \ , \quad v_l = v^0_l+ \Omega v^1_l \ , \quad u_l =
\Omega u^1_l
\ee
and similarly for the various scalar perturbation quantities. For example, in the case of the proton fraction we
have  $\delta x_\p= \sum_l \delta x_l Y_l^m$ with
\be
\delta x_l = \delta x_l^0 + \Omega \delta x_l^1
\ee

\subsection{The non-rotating problem}

At the non-rotating level the equations in Section~IV provide the following  relations
\be
\rho_\n\delta \tilde{\mu}_{\n,l}^0 + \rho_\p\delta \tilde{\mu}_{\p,l}^0 = - i \omega_0 \rho V_l^0
\label{eqone}\ee
\be
\rho_\n r\partial_r \delta \tilde{\mu}_{\n,l}^0 + \rho_\p r\partial_r  \delta \tilde{\mu}_{\p,l}^0  = -i\omega_0 \rho W_l^0
\label{eqtwo}\ee
\be
\delta \beta_l^0 = - i \omega_0 (1 -\bep) v_l^0
\ee
\be
r \partial_r \delta \beta_l^0 = - i \omega_0 (1-\bep)w_l^0
\label{db0}\ee
Meanwhile the continuity equations lead to
\be
i \omega_0 r^2 \delta \rho_l^0 + \partial_r (r \rho W_l^0) - l(l+1) \rho V_l^0 = 0
\label{contav} \ee
and
\be
i \omega_0 \rho r^2 \delta x_{l}^0 + \partial_r [ x_\p(1-x_\p)r \rho w_l^0 ] - x_\p(1-x_\p) l(l+1) \rho v_l^0 + \rho W_l^0 r \partial_r x_\p = 0
\label{contdi}\ee

Before we proceed, we will simplify the problem. Our  aim is to determine analytic approximations for the fundamental
modes of the system, including the mutual friction damping. Solving the problem numerically is, of course, straightforward but
does not lead to the same level of insight into the dependence on the various parameters. 
To facilitate an analytic solution, we will combine an incompressible background model
with compressible perturbations. This simplifies the calculations considerably.
In addition, since this is the same model that was considered by Lindblom and Mendell \cite{lm95} we can compare our final results directly
to the available literature.
We  thus assume that
$\rho_\n$ and $\rho_\p$ are both constant, while $\delta \rho_\n$ and $\delta \rho_\p$ are not.

It is also useful to introduce
a new variable for the co-moving degree of freedom. Let us define
\be
\delta h_l = { 1 \over \rho} \delta p_l =  { 1 \over \rho} ( \rho_\n\delta \tilde{\mu}_\n^l + \rho_\p\delta \tilde{\mu}_\p^l)
\ee
For a single barotropic fluid, $\delta h_l$   corresponds to the perturbed enthalpy. For a compressible background model
we would have
\be
 \rho_\n r \partial_r \delta \tilde{\mu}_\n^l + \rho_\p r \partial_r \delta \tilde{\mu}_\p^l
= \rho r \partial_r \delta h_l - \rho \delta \beta_l r \partial_r x_\p
\ee
However, for the uniform density model the gradient of the proton fraction vanishes so we simply have
\be
 \rho_\n r \partial_r \delta \tilde{\mu}_\n^l + \rho_\p r \partial_r \delta \tilde{\mu}_\p^l
= \rho r \partial_r \delta h_l
\ee
It is also worth noting that $\delta h_l$ has the same dimension as $\delta \beta_l$.

We now find that (\ref{eqone}) and (\ref{eqtwo}) can be written
\be
\delta h_l^0 = - i \omega_0  V_l^0
\ee
and
\be
r \partial_r \delta h_l^0  = -i\omega_0 W_l^0
\ee

Before we proceed, we need to decide what variables we want to work with.
 We can either remove $[\delta h_l , \delta\beta_l]$ or
$[\delta \rho_l , \delta x_\p^l]$ (or some other combination of these variables) from the problem using thermodynamic identities. Opting for the latter
possibility, we use
\be
\delta \rho_l = { \rho \over c_s^2}  \delta h_l + { \rho \alpha_1 \over c_s^2} \delta \beta_l
\ee
and
\be
\delta x_\p^l = { \alpha_1 \over c_s^2} \delta h_l
+ { \alpha_2 x_\p \over c_s^2} \delta \beta_l
\ee
In these relations we have defined, first of all, the speed of sound as
\be
c_s^2 = \left( { \partial p \over \partial \rho} \right)_\beta = \rho \left( { \partial h \over \partial \rho} \right)_\beta
\ee
We have also introduced
\be
\alpha_1 = {  c_s^2 \over \rho} \left( {\partial \rho \over \partial \beta} \right)_h
\ee
and
\be
\alpha_2 = { c_s^2  \over x_\p} \left( {\partial x_\p \over \partial \beta} \right)_h
\ee
and made use of the identity~\footnote{Ultimately, this relation follows from the fact that the partial derivatives with respect to the number densities commute, so the mixed second  derivatives of the energy functional (the ``equation of state'') must be equal.}
\be
\rho \left( { \partial x_\p \over \partial h} \right)_\beta   = \rho^2 \left( { \partial x_\p \over \partial p} \right)_\beta = \left(
{\partial \rho \over \partial \beta} \right)_p
\ee
This reduces the number of required ``thermodynamic'' quantities to three; $c_s^2$, $\alpha_1$ and $\alpha_2$.

For later convenience, it is useful to pause and consider the relative magnitude of the thermodynamic derivatives.
To do this, take as an example an overall $n=1$ polytrope with a proton fraction that is linear in the total density.
This simple model is not completely unrealistic, and moreover it is straightforward to work out all the quantities we need.
Assuming that
\be
p = K \rho^2
\ee
we find that
\be
 \left( { \partial p \over \partial \rho} \right)_\beta = 2 K \rho = c_s^2 \longrightarrow  \left( { \partial h \over \partial \rho} \right)_\beta
= { c_s^2 \over \rho}
\ee
Combine this with the assumption that the proton fraction (in equilibrium) is linear in the density.
That is, take
\be
x_\p = \alpha \left( { \rho \over \rho_c} \right)
\ee
where $\alpha\sim 10^{-1}$ and $\rho_c$ is the central density of the star. This leads to
\be
 \left( { \partial \beta \over \partial \rho} \right)_h = { 1 \over \rho} \left( { \partial h \over \partial x_\p } \right)_\beta = { 2 K \rho  \over x_\p}  = { c_s^2 \over x_\p \rho}
\ee
and
\be
 \left( { \partial \beta \over \partial x_\p} \right)_h =  { 2 K \rho \over x_\p^2 }  = { c_s^2 \over x_\p^2}
\ee
These estimates suggest that $\alpha_1 \sim \alpha_2 \sim x_\p$.
Since we expect to have $x_\p\ll 1$ it should be the case that
\be
\left( {  \partial \rho  \over \partial h} \right)_\beta \gg { 1 \over \rho}  \left( {  \partial \rho \over \partial \beta } \right)_h
\gg  \left( {  \partial x_\p  \over \partial \beta } \right)_h
\ee
This agrees with the more realistic equation of state considered by Lindblom and Mendell \cite{lm95}.
We will make explicit use of this ordering later.

Returning to the coupled system of equations, and combining
the various relations we easily arrive at the two differential equations
\be
\partial_r (r^2 \partial_r\delta h_l^0) -  l(l+1)\left[1 - {\omega_0^2 r^2 \over l(l+1) c_s^2} \right] \delta h_l^0 +
{ \omega_0^2 \alpha_1 r^2 \over c_s^2}  \delta \beta_l^0 = 0
\ee
and
\be
\partial_r (r^2 \partial_r \delta \beta_l^0)
-  l(l+1) \left[ 1 - { (1-\bep) \omega_0^2 \alpha_2 r^2 \over l(l+1) (1-x_\p) c_s^2 } \right]  \delta \beta_l^0 + {(1-\bep) \omega_0^2 \alpha_1 r^2  \over x_\p (1-x_\p) c_s^2 }  \delta h_l^0 = 0
\ee
These are the equations to be solved.

Before we proceed, it is useful to introduce dimensionless variables. First we introduce $\omega_0=\sigma_0 \bar{\omega}$ where
$\bar{\omega}^2=GM/R^3$.
Then we consider a new radial variable
\be
s = {\bar{\omega} r \over c_s}
\ee
This means that have
\be
\partial_s (s^2 \partial_s \delta h_l^0)  -  l(l+1)\left[1 - {\sigma_0^2 s^2 \over l(l+1)} \right] \delta h_l^0 +  \alpha_1 \sigma_0^2 s^2  \delta \beta_l^0 = 0
\label{eqh0}\ee
together with
\be
\partial_s (s^2 \partial_s \delta \beta_l^0)
- l(l+1) \left[ 1 - { (1-\bep) \alpha_2 \sigma_0^2 s^2 \over l(l+1) }\right]  \delta \beta_l^0 + {(1-\bep) \alpha_1  \sigma_0^2 s^2 \over x_\p}   \delta h_l^0 = 0
\label{eqb0}\ee
For simplicity, we have assumed that $x_\p$ is small ($\ll 1$). This should always be the case.
When the equations are written in this form it becomes apparent that the coupling term in
(\ref{eqb0}) is more important than that in (\ref{eqh0}). From this one can deduce that there should exist solutions to
the problem such that $\delta h_l^0 \gg \delta \beta_l^0$. These are the modes that we will focus on. This is natural, if our main focus
is on oscillations that  radiate gravitational waves at a significant level, e.g. by being driven unstable \cite{nareview}.

As already mentioned, since we have assumed that the background configuration is uniform, our model
is identical to the incompressible/compressible model considered by Lindblom and Mendell \cite{lm95}.
From their work we know that we can write down the solution to the coupled equations in closed form using
(spherical) Bessel functions. This solution would contain all the modes of the system, fundamental modes
and pressure modes with varying degree of co- and counter-moving character. However,
this solution is not very practical for our present purposes. If we want to solve the order $\Omega$ problem
explicitly, rather than estimate the mutual friction damping via the energy integral approach (as Lindblom and Mendell did),
we need to be able to solve another system of equations where the leading order mode-solution acts as source.
Expressed in terms of the Bessel-function solutions, the order $\Omega$ problem is very messy.
Hence, we  opt for a different strategy and introduce yet another simplifying approximation.

In order to proceed analytically, let us assume that  $s^2  \ll 1 $,
in which case we can attempt to solve the problem using a power series.
Is this reasonable? Well, let us again consider the case of an $n=1$ polytrope.
In that case 
\be
K = 2 \pi G \left( {R \over \pi} \right)^2
\ee
and it follows that
\be
s^2 \le { \pi^2 \over 3 } \left( { \bar{\rho} \over \rho} \right)
\ee
where $\bar{\rho}$ is the average density of the star. This shows that, in our uniform parameter model, the power series
Ansatz makes sense as long as we assume $\rho \gg 3 \bar{\rho}$. That is, the calculation should be relevant for the conditions in a neutron star core.
However, it is obviously not completely consistent. The assumptions will  not hold
 near the surface of the star,  since
one tends to have $c_s^2 \to 0$ as $r\to R$.
However, since the surface region is already dealt with in a rather ad hoc way this does not  concern us too much.

Now that we have a small parameter in the problem, we can try to find a power series solution.
It is natural to first rewrite the coupled problem as a single fourth-order equation for (say)
$\delta h_l^0$. This is easily done by combining (\ref{eqh0}) and (\ref{eqb0}). Making the Ansatz
\footnote{Note that we are neglecting the entrainment in these equations. This is, however, not important.
By including the relevant entrainment factors one can show that they do not affect the f-mode result to
the order of approximation at which our solution is valid.}
\be
\delta h_l^0 = s^l \sum_{n=0}^N a_n s^n
\ee
we find that the first few coefficients are determined by
\be
a_1=a_3=a_5 = 0
\ee
and
\be
{ \sigma_0^2(\alpha_1^2 - \alpha_2 x_\p) \over x_\p } a_0 - 2( 1+ \alpha_2)(2l+3)a_2 - { 8 (2l+3)(2l+5) \over \sigma_0^2} a_4 = 0
\label{rec1}\ee

We now insert this solution into (\ref{eqh0}). If we write
\be
\delta \beta_l^0 = s^l \sum_{n=0}^N b_n s^n
\ee
then  we must have
\be
b_0 = - { \sigma_0^2 a_0 + 2(2l+3)a_2 \over \alpha_1 \sigma_0^2}
\ee
and
\be
b_2 = - { \sigma_0^2 a_2 + 4(2l+5)a_4 \over \alpha_1 \sigma_0^2}
\ee

To complete the solution, we need to satisfy the boundary conditions. We want
\be
\partial_s \delta \beta_l^0 = 0 \qquad \mbox{ at} \qquad s = {\bar{\omega} R \over c_s}
\ee
Keeping the first two terms in the series for $\delta \beta_l^0$, this condition leads to another
relation between the three coefficients $a_0$, $a_2$ and $a_4$. Combining this relation with \eqref{rec1} we
arrive at an expansion for $\delta h_l^0$ where the overall scaling is given by $a_0$, and the only other
unknown parameter is the frequency $\sigma_0$. To fix the frequency, we impose the remaining boundary condition. That is,
we require
\be
\partial_r \delta h_l^0 + \left( { \omega_0^2 \over \partial_r \tilde{\mu} } \right) \delta h_l^0 = 0  \qquad \mbox{ at} \qquad r=R
\ee
where (for a uniform background model)
\be
\left.\partial_r \tilde{\mu}\right|_{r=R} = - { 4 \pi G\rho R \over 3} = - \bar{\omega}^2 R
\ee
This leads to the condition
\be
\partial_s \delta h_l^0 - { c_s \over \bar{\omega} R} \sigma_0^2 \delta h_l^0 = 0 \qquad \mbox{ at} \qquad s = {\bar{\omega} R \over c_s}
\ee
Some algebra now leads to a solution with frequency
\be
\sigma_0^2 \approx l \left[ 1 - { 1 \over 2l+3} \left( { \bar{\omega} R \over c_s } \right)^2 \right]
\ee
That is, we have
\be
\omega_0^2 \approx { l GM \over R^3} \left[ 1 - { 1 \over 2l+3} \left( { \bar{\omega} R \over c_s } \right)^2 \right]
\ee
The leading order result is exactly what one would expect for an incompressible fluid ball in the Cowling approximation.
The first correction to this is directly associated with the compressibility. The presence of the second fluid degree of freedom, e.g. the link to
$\delta \beta$, appears at the next order of approximation. It is also worth pointing out that 
 the solution is such that (omitting geometric factors of $l$)
\be
b_0 \sim { \alpha_1 \over x_\p} \left( { \bar{\omega} R \over c_s } \right)^2  a_0
\label{B0}\ee
This demonstrates that  the mode  we have determined is such that $\delta h_l^0 \gg \delta \beta_l^0$.
The associated fluid motion is, indeed,
predominantly co-moving.  The conclusion that the co-moving f-mode is hardly at all affected by the two-fluid nature of the system 
accords well with the results of Lindblom and Mendell \cite{lm95}.

\subsection{The slow-rotation corrections}

Having approximated
the f-mode solution to the non-rotating problem, we will now work out the first order slow-rotation corrections. This will include the mutual friction damping.

The equations that need to be solved at order $\Omega$ are, first of all
\be
2(l+2)(W_l^0-lV_l^0)Q_{l+1} Y_{l+1}^m - 2(l-1)[W_l^0+(l+1)V_l^0]Q_l Y_{l-1}^m
+l(l+1)\omega_0 U_l^1 Y_l^m = 0
\ee
which determines the axial rotational correction $U_l^1$ to the f-mode. We are not going to determine this quantity here
because it does not affect the mode damping, which is our main concern.
A similar equation for the counter-moving degree of freedom  determines the axial  correction, $u_l^1$.
This is also not of immediate relevance for our discussion, so we do not consider it.

From Section~IV we see that the equations we actually need to solve are
\be
[l(l+1)\omega_1-2m] V_l^0+l(l+1)\omega_0 V_l^1 - 2m W_l^0   - il(l+1)\delta h_l^1 = 0
\ee
and
\be
i r \partial_r \delta h_l^1 + 2mV_l^0-\omega_1 W_l^0 - \omega_0 W_l^1 = 0
\ee
together with
\begin{multline}
- \Big\{ il(l+1) \delta \beta_l^1 + 2 m \bBp w_l^0-[l(l+1)\omega_1(1-\bep)
-2m\bBp]v_l^0 -l(l+1)\omega_0(1-\bep)v_l^1 \\
+2i\bB[1-(l+3)Q_{l+1}^2+(l-2)Q_l^2]w_l^0
+2i\bB[m^2+l(l+3)Q_{l+1}^2+(l+1)(l-2)Q_l^2]v_l^0\Big\}Y_l^m\\
-2i(l+3)(l v_l^0-w_l^0)\bB Q_{l+1}Q_{l+2} Y_{l+2}^m
- 2i(l-2)[w_l^0-(l+1)v_l^0]\bB Q_l Q_{l-1} Y_{l-2}^m = 0
\end{multline}
and
\begin{multline}
\Big\{ ir\partial_r \delta \beta_l^1 - [\omega_1(1-\bep)-2i\bB(1-Q_l^2-Q_{l+1}^2] w_l^0
-\omega_0(1-\bep)w_l^1 + 2[m\bBp-i((l+1)Q_l^2-lQ_{l+1}^2)\bB]v_l^0\Big\}Y_l^m \\
+2i(lv_l^0-w_l^0)\bB Q_{l+1} Q_{l+2}Y_{l+2}^m - 2i [ w_l^0+(l+1)v_l^0] \bB Q_l Q_{l-1}
Y_{l-2}^m = 0
\end{multline}

If we want to determine the rotational correction to the f-mode, then we only need to consider the
order $\Omega$ terms that are sourced by non-rotating terms. Thus the problem reduces to solving
\be
l(l+1)[i\delta h_l^1-\omega_0 V_l^1] = [l(l+1)\omega_1-2m]V_l^0-2mW_l^0
\ee
and
\be
ir\partial_r \delta h_l^1 - \omega_0 W_l^1 = \omega_1 W_l^0 - 2mV_l^0
\ee
together with the continuity equation (assuming a uniform background)
\be
i \omega_0 r^2 \delta \rho_l^1 +  \rho \partial_r (r W_l^1) - l(l+1) \rho V_l^1 =
- i \omega_1 r^2 \delta \rho_l^0
\ee
Note that there is no multipole coupling in these equations.
For the other degree of freedom we have
\begin{multline}
l(l+1)[i \delta \beta_l^1 - \omega_0(1-\bep)v_l^1] = - 2m\bBp w_l^0 +   [l(l+1)\omega_1 (1-\bep) -2m\bBp] v_l^0   \\
- 2i\bB[1-(l+3)Q_{l+1}^2+(l-2)Q_l^2] w_l^0
- 2i\bB[m^2+l(l+3)Q_{l+1}^2+(l+1)(l-2)Q_l^2]v_l^0
\end{multline}
and
\be
i r \partial_r \delta \beta_l^1 - \omega_0(1-\bep)w_l^1 = [ \omega_1(1-\bep)-2i\bB(1-Q_l^2-Q_{l+1}^2)] w_l^0 
- 2\{ m\bBp - i [(l+1)Q_l^2 - lQ_{l+1}^2]\bB\} v_l^0
\ee
together with
\be
i \omega_0 r^2 \delta x_l^1 + x_\p(1-x_\p) [ \partial_r (r w_l^1) - l(l+1) v_l^1] =
-i \omega_1 r^2 \delta x_{l}^0
\ee
We want to find solutions that satisfy the boundary conditions
\be
i \omega_0 \delta h_l^1 + { 1 \over r} W_l^1 \partial_r \tilde{\mu} = - i \omega_1 \delta h_l^0 \ , \qquad \mbox{ at } r = R
\ee
and
\be
i \omega_0 \partial_r \delta \beta_l^1 = - i \omega_1 \partial_r \delta \beta_l^0 = 0 \ , \qquad \mbox{ at } r = R
\ee

After some manipulations (making use of the leading order relations) we arrive at the two coupled equations
\be
\partial_r (r^2 \partial_r\delta h_l^1) -  l(l+1)\left[1 - {\omega_0^2 r^2 \over l(l+1) c_s^2} \right] \delta h_l^1 +
{ \omega_0^2 \alpha_1 r^2 \over c_s^2}  \delta \beta_l^1 =
 - {  2 \omega_1 \omega_0 r^2 \over c_s^2 } \left[ \delta h_l^0  + \alpha_1 \delta \beta_l^0 \right]
\ee
and
\begin{multline}
\partial_r (r^2 \partial_r \delta \beta_l^1)
-  l(l+1) \left[ 1 - { (1-\bep) \omega_0^2 \alpha_2 r^2 \over l(l+1) (1-x_\p) c_s^2 }\right]  \delta \beta_l^1 +
 { \omega_0^2 r^2 (1-\bep) \alpha_1   \over x_\p (1-x_\p) c_s^2 } \delta h_l^1 = \\
= - { 2 \omega_0 \alpha_1 r^2 \over x_\p (1-x_\p) c_s^2 } [ (1-\bep)\omega_1 -i\bB(1-Q_l^2-Q_{l+1}^2)]  \delta h_l^0 \\
+
{ 2i\bB \over (1 -\bep)\omega_0 } [ 1 + (2l-1)Q_l^2 - (2l+3)Q_{l+1}^2]   r \partial_r \delta \beta_l^0 \\
+ {2i \bB \over (1- \bep) \omega_0} [ m^2 - l(l+1) + (l+1)(2l-1)Q_l^2 + l (2l+3)Q_{l+1}^2] \delta \beta_l^0 \\
- {2 \alpha_2  \omega_0 r^2 \over (1-x_\p) c_s^2} [ (1-\bep) \omega_1 - i (1-Q_l^2-Q_{l+1}^2) \bB] \delta \beta_l^0
\label{db2}\end{multline}

The problem has the anticipated form, a coupled system of equations for $\delta h_l^1$ and $\delta \beta_l^1$ which differs
from the non-rotating problem only by the presence of leading order source terms. To proceed, we will follow the same strategy as in the leading order
calculation. However, it is beneficial to first note that the source term in the second equation simplifies considerably if we focus on the
$l=m$ f-modes. Since $Q_{m+1}^2 = 1/(2m+3)$ we see that the factors in front of the first two $\delta \beta_l^0$ pieces in the
right-hand side of \eqref{db2} are then identically zero. Moreover, we know from the leading order calculation that $\delta \beta_l^0\ll \delta h_l^0$, cf. \eqref{B0}.
As long as we are only interested in the leading order rotational correction and the leading order mode damping,
this allows us to neglect also the remaining $\delta \beta_l^0$ part of the source. Using $\omega_0 r/c_s = \sigma_0 s$ as well as
$\sigma_1 = \omega_1/\omega_0$ and assuming that $x_\p\ll 1$ we then arrive at the simplified equations
\be
\partial_s (s^2 \partial_s\delta h_l^1) -  l(l+1)\left[1 - {\sigma_0^2 s^2 \over l(l+1)} \right] \delta h_l^1 +
 \sigma_0^2 \alpha_1 s^2   \delta \beta_l^1 =
 -  2 \sigma_0^2 \sigma_1 s^2  \delta h_l^0
\label{eqh1}\ee
and
\be
\partial_s (s^2 \partial_s \delta \beta_l^1)
-  l(l+1) \left[ 1 - { (1-\bep) \sigma_0^2 \alpha_2 s^2 \over l(l+1) }\right]  \delta \beta_l^1 +
 { \sigma_0^2 s^2 (1-\bep) \alpha_1   \over x_\p} \delta h_l^1 
= - { 2  \alpha_1 s^2 \mathcal{D} \over x_\p }  \delta h_l^0
\label{eqb1}\ee
where
\be
\mathcal{D} =  \sigma_0^2 \left[ (1-\bep) \sigma_1 - { 2 i(l+1) \bB \over \sigma_0 \bar{\omega} (2l+3)}\right]
\ee

Combining the two equations, making the same Ansatz as in the non-rotating case
\be
\delta h_l^1 = s^l \sum_{n=0}^N a_n s^n
\ee
and taking the source term to be
$\delta h_l^0 = C s^l$ \footnote{Since we are only interested in the leading corrections it is sufficient to use the leading term in the non-rotating solution as source.}, we find that
\be
a_1=a_3=a_5 = 0
\ee
Meanwhile, we have
\be
{ \sigma_0^2(\alpha_1^2 - \alpha_2 x_\p) \over x_\p } a_0 - 2( 1+ \alpha_2)(2l+3)a_2 -
{ 8 (2l+3)(2l+5) \over \sigma_0^2} a_4 + 2 \left( { \alpha_1^2 \mathcal{D}\over x_\p} - \sigma_0 \sigma_1 \alpha_2  \right) C = 0
\label{rec2}\ee
The difference now is that we are only interested in the particular solution due to the presence of the
source term. In order to remove the unwanted homogeneous solution we set $a_0=0$.
Then \eqref{rec2} becomes a relation between the known mode amplitude $C$ and the two coefficients $a_2$ and $a_4$.
As in the non-rotating problem, we get a second such relation from the boundary condition for $\delta \beta_l^1$.
From  (\ref{eqh1}) and
\be
\delta \beta_l^1 = s^l \sum_{n=0}^N b_n s^n
\ee
we get
\be
b_0 = - { 2 \over \alpha_1} \left[ { (2l+3) \over \sigma_0^2} a_2 + \sigma_1  C  \right]
\ee
and
\be
b_2 =   - { 1 \over \alpha_1} \left[ a_2 + {4(2l+5)a_4 \over  \sigma_0^2} \right]
\ee
This solution has to satisfy the surface condition

\be
\partial_s \delta \beta_l^1 = 0 \qquad \mbox{ at} \qquad s = {\bar{\omega} R \over c_s}
\ee
This leads to another relation between $C$, $a_2$ and $a_4$. This means that we can write down
the solution for $\delta h_l^1$ with an overall scale $C$ (as expected for the particular solution) and
with the frequency correction $\sigma_1$ as the only remaining undetermined quantity.

The final condition to be satisfied can be written (again, for $l=m$)
\be
\partial_s \delta h_l^1 - { c_s \over \bar{\omega} R} \sigma_0^2 \delta h_l^1 =  { c_s \over \bar{\omega} R}
\left[ \sigma_1 \left( \sigma_0^2 +  l  \right) - {2l \over \bar{\omega} \sigma_0} \right] C s^l
\qquad \mbox{ at} \qquad s = {\bar{\omega} R \over c_s}
\ee

Inserting the power series solution, one can show that this condition leads to the leading order rotation correction of the f-mode frequency
being
\be
\mathrm{Re}\ \sigma_1 \approx {1 \over \bar{\omega} l^{1/2}}
\ee
Meanwhile the leading damping term (the imaginary part of $\sigma_1$) is
\be
\mathrm{Im}\ \sigma_1 \approx   { 2 (l+1) (3l+5)  \over \sqrt{l} (2l+3)^3 (2l+5) } { \alpha_1^2 \over x_\p}  \bB \left( {\bar{\omega}^3 R^4 \over  c_s^4} \right)
\ee
These are the final results of the f-mode analysis. After retracing our steps  to recall the various definitions, 
we find that  the mutual friction damping follows from
\be
\mathrm{Im}\ \omega_1 \approx { 2 (l+1) (3l+5) \over (2l+3)^3 (2l+5)} {\alpha_1^2  \over x_\p^2} \left({GM \over R c_s^2 } \right)^2 \mathcal{B} = { 2 (l+1) (3l+5) \over (2l+3)^3 (2l+5)} { 1 \over \rho_\p^2}
\left({ \partial \rho \over \partial \beta} \right)_h^2
\left({GM \over R }\right)^2 \mathcal{B}
\ee
That is, the damping timescale is
\be
\tau = { 1 \over \mathrm{Im}\ \omega_1 \Omega} = { (2l+3)^3 (2l+5) \over 2 (l+1) (3l+5)}\left[  { 1 \over \rho_\p^2}
\left({ \partial \rho \over \partial \beta} \right)_h^2
\left({GM \over R }\right)^2 \right]^{-1} { 1 \over \mathcal{B} \Omega} 
\label{mfdamp}\ee
This result completes our analysis of the (co-moving) f-mode in a superfluid neutron star.

It is obviously relevant to compare the estimated f-mode damping timescale to previous work.
To do this,  we first need to recall that all previous work has focused on the case where electron scattering off the
vortex array is the main cause of mutual friction. Then we have \cite{mendell,trev1}
\be
\mathcal{B} \approx 4 \times 10^{-4} \left({ m_\p - m_\p^* \over m_\p} \right)^2 \left( {m_\p \over m_\p^* } \right)^{1/2}
\left( {x_\p \over 0.05} \right)^{7/6} \left( {\rho \over 10^{14} \mathrm{g/cm}^3 } \right)^{1/6}
\label{Bcanon}\ee
where we have used the relation between the entrainment and the effective proton mass;
\be
\varepsilon_\p = 1 - { m_\p^* \over m_\p}
\label{effp}\ee
Taking $m_\p^*/m_\p=0.3$ we  have $\mathcal{B} \approx 5.5\times10^{-4}$ in good agreement with the 
result used by Lindblom and Mendell in their investigation of the f-mode problem \cite{lm95}.   
The overall scaling with $l$ in (\ref{mfdamp})  also appears to be similar to their result. This is evident from the results in Table~\ref{damptab}
which compares our results to data
from Table~1 in \cite{lm95}. This comparison shows that, in the  $m_\p^*/m_\p=0.3$ case, our damping times 
are about a factor of 2 longer than those estimated by Lindblom and Mendell. 
The results also differ in the predicted dependence on the entrainment.
In our calculation, the entrainment only enters \eqref{mfdamp} indirectly through its effect on $\mathcal{B}$.
The data given by Lindblom and Mendell hints at a different behaviour. In the $m_\p^*/m_\p=0.8$ case we find that the difference between  our damping result and
 Table~1 in \cite{lm95} is closer to a 
factor of 3.

Most likely the main difference originates from the use of the energy integral approach in one case and
the direct determination of dissipative mode solutions in the other.
In order to check this assumption, we have  estimated the mutual friction damping using the
energy integral approach (together with our leading order f-mode solution). That is, we evaluate (\ref{tauest}) for the
non-rotating f-mode solution. Then we find that 
\be
\tau = { (2l+3)^3 (2l+5) \over 6(2l^2+6l+5) }\left[  { 1 \over \rho_\p^2}
\left({ \partial \rho \over \partial \beta} \right)_h^2
\left({GM \over R }\right)^2 \right]^{-1} { 1 \over \mathcal{B} \Omega} 
\label{intdamp1}\ee
This scaling with $l$ here   differs somewhat from  that in (\ref{mfdamp}).
This introduces a numerical factor of $\approx 2$ for small values of $l$. This factor brings our results very close to those of Lindblom and Mendell, cf. Table~\ref{damptab}. 
The main difference between the energy integral result and the full dissipative mode calculation is a geometric factor of order unity. 
This is what one would expect. At the end of the day the astrophysical implications of the results are the same.  

\begin{table}
\caption{Estimated mutual friction damping timescales for the (co-moving) f-mode of a superfluid neutron star. We compare our results to previous work by Lindblom and Mendell~\cite{lm95}. 
The parameters are those discussed in the main text, and correspond to  $m_\p^*/m_\p=0.3$. The stellar model has radius 15~km and average density $4\times10^{14}$~g/cm$^3$.
Note that, in our case the non-rotating f-mode frequency is given by $\omega_0^2 = 4l\Omega_0^2/3$, where $\Omega_0 = \sqrt{\pi G \bar{\rho}}$ and $\bar{\rho}$ is the average density. 
The difference between the listed frequencies (15\% for $l=2$ and decreasing with increasing $l$) 
should mainly be due to our use of the Cowling approximation. The dissipative mode result \eqref{mfdamp} for the mutual friction 
timescales (which scales as $\Omega_0/\Omega$) differs by about a factor of 2 from the tabulated Lindblom-Mendell results. By comparing to \eqref{intdamp1} we learn that this is mainly due
to different geometric factors in the energy integral approach and the direct dissipative mode calculation.
}
\begin{tabular}{|c|cc|ccc |}
\hline 
 & \multicolumn{2}{c}{Lindblom-Mendell} & \multicolumn{3}{c}{This work}   \\
\hline
\hskip0.25cm $l$ \hskip0.25cm & \hskip0.5cm $\omega(0)/\Omega_0$ \hskip0.5cm & \hskip0.5cm  $\tau \Omega_0$ \hskip0.5cm & \hskip0.5cm $\omega(0)/\Omega_0$  \hskip0.5cm &  \hskip0.5cm $\tau\Omega_0$ from \eqref{intdamp1}  \hskip0.5cm 
&  \hskip0.5cm  $\tau\Omega_0$ from \eqref{mfdamp} \hskip0.5cm\\
\hline
2 & 1.407 & $1.071\times10^4$  & 1.63  & $9.2\times10^3$  & $2.1\times10^4$ \\
3 & 1.809 & $1.638\times10^4$  & 2     & $1.5\times10^4$  & $3.2\times10^4$\\
4 & 2.141 & $2.347\times10^4$  & 2.31  & $2.1\times10^4$  & $4.6\times10^4$\\
5 & 2.430 & $3.195\times10^4$  & 2.58  & $2.9\times10^4$  & $6.1\times10^4$\\
6 & 2.689 & $4.182\times10^4$  & 2.83  & $3.8\times10^4$  & $8.0\times10^4$\\
\hline
\end{tabular}
\label{damptab}
\end{table}

Before we move on, it is worth making a comment on the apparent lack of entrainment scaling. The damping timescale in (\ref{mfdamp}) does not depend directly on the entrainment parameter $\varepsilon$ due to a series of
cancellations that occur when we impose the order $\Omega$ boundary condition $\nabla_i \delta\beta=0$. If we were to impose the (slightly more realistic) condition $\delta w^r=0$
at the surface (corresponding to a common surface), these cancellations would not occur and there would be an explicit dependence on $\varepsilon$ in the damping timescale. This does not, however, 
affect the numerical results significantly.

\subsection{The f-mode instability window}

\begin{figure}[t]
\centering
\includegraphics[width=10cm,clip]{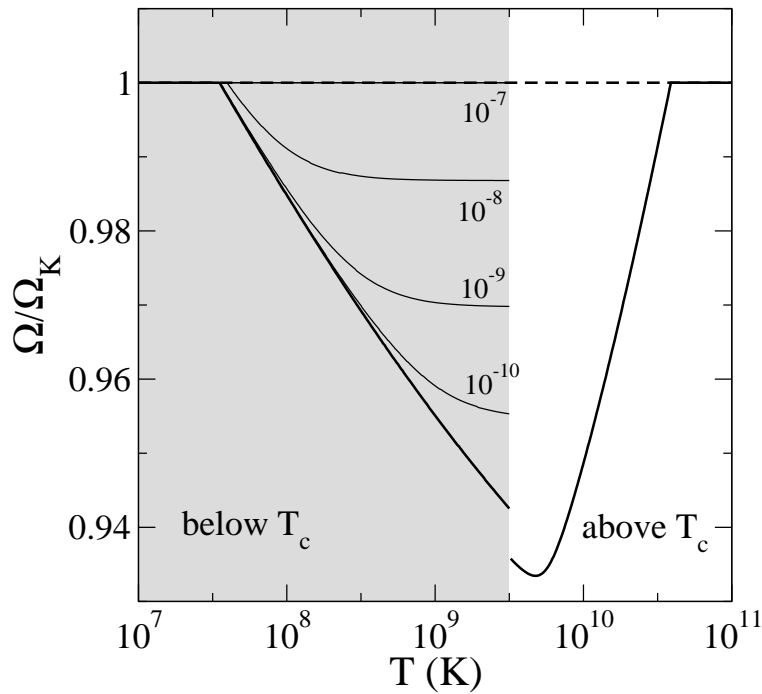}
\caption{The f-mode instability window for the $l=m=4$ f-mode. The data for gravitational radiation reaction, shear- and bulk viscosity are taken from \cite{il91}. The instability is active above 
a critical rotation rate (thick solid line) at each given temperature. The mutual friction, which acts only below the superfluid transition temperature (here taken to be 
$5\times10^9$~K, lower temperatures are indicated by the grey region in the figure) is estimated using \eqref{mfdamp} and the parameter values given in the main text. To illustrate the role of a weak mutual friction we show (as thin solid lines) the instability curves for $\mathcal B$ in the range $10^{-10}-10^{-7}$, more than three orders of magnitude weaker than the canonical value \eqref{Bcanon}. The solid rotation rate of the star is given as a fraction of the breakup rate $\Omega_K \approx 0.639 \Omega_0$.  }
\label{fwindow}
\end{figure}

To conclude the discussion of the superfluid f-mode, it is worth considering the impact of the results on the 
gravitational-wave driven instability of this mode. The close agreement between our mutual friction damping rates and 
the results of Lindblom and Mendell \cite{lm95} obviously means that their key conclusions stand. 
That is, the instability of the f-mode is likely to be completely suppressed in a superfluid neutron star. However, we think that
this result has sometimes  been misunderstood. The result does not show that the secular f-mode instability cannot play a role for astrophysical neutron stars. To show 
this, we have combined the different timescales for gravitational-wave growth of an unstable mode with the damping due to shear- and bulk viscosity from 
Ipser and Lindblom \cite{il91}. The results, for the $l=m=4$ f-mode that leads to the strongest instability in a Newtonian star, 
are shown in Figure~\ref{fwindow}. The data in the figure corresponds to an $n=1$ polytrope with mass $1.5 M_\odot$ and 
12.533~km (the average density is $3.6\times10^{14}$ g/cm$^3$). In order to connect this with our mutual friction approximation, we have used the model parameters from \cite{lm95}, i.e., 
\begin{eqnarray*}
\rho &=& 4\times 10^{14} \mathrm{g/cm}^3 \\
x_\p &=& 0.06 \\
\left( {\partial \rho \over \partial \beta} \right)_p &=& 1.911 \times 10^{-7} \mathrm{gs}^2/\mathrm{cm}^5
\end{eqnarray*}
Combined with the canonical value for $\mathcal B$ given in \eqref{Bcanon} these parameter values lead to the 
mutual friction damping completely overwhelming any gravitational-wave driving of the f-mode. Table~\ref{damptab} provides similar results  for the model considered in \cite{lm95}, a star with radius 15~km and 
average density $4\times10^{14}$~g/cm$^3$ (which means that the mass is quite large, $2.84 M_\odot$). We have scaled the frequencies
and the timescales using $\Omega_0 = \sqrt{\pi G \bar{\rho}}$ where $\bar{\rho}$ is the average density.

However, this result is
only relevant below the critical temperature at which the stars core become superfluid. Suppose that we take the critical temperature to be $5\times10^9$~K, which is a typical value \cite{npa}. 
Then the f-mode instability window remains unaltered in hotter stars. This is evident from Figure~\ref{fwindow}. Of course, as soon as a sizeable part of the core is superfluid, 
the instability is no longer present. However, it seems that there is still scope for the 
unstable f-modes to play a role in the evolution of a nascent neutron star born spinning near the breakup velocity.
One should also remember that the instability is stronger in a relativistic model. In fact, in this case the 
$l=m=2$ mode may also become unstable. Based on the available evidence it would be premature to rule out the 
f-mode instability for realistic neutron star models. The problem requires further attention.  

In Figure~\ref{fwindow} we also show the effect of a weaker mutual friction. Suppose that \eqref{Bcanon} is, for some reason, not the typical value. Our understanding of neutron star core physics is not complete, so it is interesting to consider a range of possibilities. The different thin solid curves in the figure show the effect of mutual friction 
for the given values of $\mathcal{B}$. For the considered $1.5M_\odot$ model the results correspond to (for $l=4$)
\be
\tau \Omega_0 \approx 2.5\times10^5 \left( { \Omega \over \Omega_0} \right)^{-1} \left( { \mathcal{B} \over 10^{-4} } \right)^{-1} 
\ee

From these results we learn that the mutual friction must be at least three orders 
of magnitude weaker than the canonical value in order for the f-mode to be unstable  below the superfluid transition temperature. 

Finally, it is worth noting the following. The coefficient $\mathcal{B}'$ was not present in our final f-mode 
equations. This is due to a series of, perhaps surprising, cancelations. The upshot of this is that our results are
also valid in the strong drag regime. Expressed in terms of the drag parameter $\mathcal R$, the data in Figure~\ref{fwindow} 
shows that the mutual friction suppresses the f-mode instability in a superfluid neutron star with 
$10^{-7} < \mathcal{R} < 10^7$. This conclusion is interesting since the strong drag regime has not been considered before. 
It also shows that the suppression of the f-mode takes place for much of the plausible parameter range.

\section{The r-modes}

Having explored the f-mode in a superfluid star, revisiting the issue of the mutual friction damping, we will now
consider the Coriolis driven r-modes in a slowly rotating star. The r-modes are interesting because they
 also suffer a gravitational-wave driven instability \cite{nareview}.
In contrast to the f-modes, which only become unstable
at fast rotation rates, the r-mode instability may set in already at quite modest spins.

In a single fluid star, the r-modes are purely axial to leading order.
Moreover, their frequency is linear in the rotation rate.
Hence, it is natural to make the Ansatz
\be
\omega = \omega_0 \Omega
\ee
together with
\be
W_l = \Omega W^1_l \ , \quad V_l = \Omega V^1_l \ , \quad U_l = U^0_l
\ee
and
\be
w_l = \Omega w^1_l \ , \quad v_l = \Omega v^1_l \ , \quad u_l =
\Omega u^0_l
\ee
Note that, if we want to work out the order $\Omega$ corrections to the mode (with the above ordering)
we will first need to
account for the centrifugal force and the change in shape of the star. We will discuss this problem elsewhere \cite{bryn}.
Here we will focus on the problem at linear slow-rotation order.

From the general slow-rotation equations in Section~4, we immediately see that the r-mode assumption
decouples the two degrees of freedom. First of all, the average vorticity equation leads to
\be
\left[ l(l+1) \omega_0 - 2m \right] U_l^0 Y_l^m = 0
\ee
This shows that we must have a single multipole solution, with frequency
\be
\omega_0 = { 2m\over l(l+1)}
\label{rnormal}\ee
To determine the associated eigenfunction we consider the divergence equation and the radial Euler
equation. These lead to the recurrence relations
\be
-i n(n+1) \delta h_n^1 - 2(n-1)(n+1) Q_n U_{n-1}^0 - 2[(n+1)^2-1] Q_{n+1} U_{n+1}^0 = 0
\ee
and
\be
ir \partial_r \delta h_n^1 + 2(n-1) Q_n U_{n-1}^0 - 2 (n+2) Q_{n+1} U_{n+1}^0 = 0
\ee
For simplicity, we have assumed that the background is uniform (as in the f-mode analysis in the previous section).
These equations show that the only way to have a single multipole axial solution is to have
$U_m^0 \neq 0$. This follows since $Q_m=0$. In other words, we will have non-trivial modes only for $l=m$. 
Inserting $n=m-1$ in the two equations we have
\be
-i (m+1) \delta h_{m+1}^1 - 2m Q_{m+1} U_m^0 =0
\ee
and
\be
i r \partial_r \delta h_{m+1}^1 + 2mQ_{m+1} U_m^0 = 0
\ee
These combine to
\be
r \partial_r U_m^0 - (m+1) U_m^0= 0
\ee
and the familiar solution
\be
U_m^0 = A r^{m+1}
\ee
This analysis shows that, to leading order, the standard r-mode remains unchanged in a superfluid star.
We also see that we need to go to higher orders in rotation if
we want to determine the mutual friction damping of these modes.
Such calculations have been
carried out by Lindblom and Mendell \cite{lm00} and Lee and Yoshida \cite{yl1}. Motivated by the 
recent evidence that the strong drag regime may be relevant, we are currently revisiting this problem \cite{bryn}.

Now consider the counter-moving degree of freedom. In that case, the difference vorticity equation leads to
\be
\left\{ l(l+1) (1 - \bep) \omega_0 - 2m \bBp - 2i [l(l+1)-m^2 ] \bB\right\} u_l^0 Y_l^m = 0
\ee
That is, there should exist a single multipole solution with frequency
\be
\omega_0 = { 1 \over 1- \bep} \left\{ { 2 m \over l(l+1)} \bBp + \frac{2i}{l(l+1)} [ l(l+1) - m^2] \bB \right\}
\label{rcounter}\ee
As in the co-moving problem, the associated eigenfunctions follow from the divergence equation
and the radial Euler equations. These lead to the recurrence relations;
\be
- i n(n+1) \delta \beta_n^1 - 2(n+1) [(n-1) \bBp +im\bB] Q_n u_{n-1}^0 - 2n[(n+2)\bBp -im\bB] Q_{n+1} u_{n+1}^0 = 0
\ee
and
\be
i r \partial_r \delta \beta_n^1  + 2 [ (n-1)\bBp + im \bB] Q_n u_{n-1}^0
- 2 [(n+2)\bBp - im \bB] Q_{n+1} u_{n+1}^0 =0
\ee
Again, it is easy to see that the only way to have a single multipole solution is to have $u_m^0 \neq 0$, i.e. 
we must have $l=m$.
This leads to
\be
-i (m+1) \delta \beta_{m+1}^1 - 2 m [\bBp + i\bB] Q_{m+1} u_m^0 = 0
\ee
and
\be
i r \partial_r \delta \beta_{m+1}^1 + 2m [ \bBp + i \bB] Q_{m+1} u_m^0 = 0
\ee
That is, we have
\be
r \partial_r u_m^0 - (m+1) u_m^0  = 0
\ee
which means that the counter-moving solution also takes the form
\be
u_m^0  = B r^{m+1}
\ee

What do we learn from this exercise? First of all, we should recognize that we have been somewhat cavalier in the
discussion. Since we have assumed that $\bep$, $\bBp$ and $\bB$ are all constant,
the analysis leading to \eqref{rcounter}
is clearly only valid for uniform background models. This tells us that the purely axial
counter-moving solution only exists for this simplified model. In a more general case, this mode will become
an axial-led inertial mode. In the weak drag regime, these inertial modes have been determined numerically by  Lee and Yoshida \cite{yl1}.

The counter-moving r-modes
are nevertheless interesting. Two particular features are worth noting. Let us first consider the mode pattern speed
\be
\sigma_p = - {\mathrm{Re}\ \omega\over m}
\ee
In the case of the normal r-mode we see from \eqref{rnormal} that the pattern speed is always negative. That is, these modes
are retrograde with respect to the star's rotation. This is the criterion that renders the mode unstable
to gravitational-wave emission at all rotation rates (in an otherwise non-dissipative star).
Meanwhile, from \eqref{rcounter} we find that
\be
\sigma_p =  - { 1 \over 1- \bep}  { 2 m \over (m+1)} \bBp = - { 1 \over 1- \varepsilon_\n/x_\p }  { 2 m \over (m+1)} \left[ 1 - { \mathcal{B}' \over x_\p} \right]
\ee
This relation shows that the second class of superfluid r-modes may, in fact, be prograde. For this to be the case we must have
(assuming that $1 - \bep >0$ , see below)
\be
\mathcal{B}' > x_\p
\ee
which may well happen. Recall that the mutual friction parameter is related to the induced friction on the vortex. From \eqref{bvsr} we see that
we need to have
\be
\mathcal{R}^2 > { 1 \over 1 - x_\p} > 1
\ee
in order for the mode to be prograde. Clearly, systems in the strong drag regime (where $\mathcal{R} \to \infty$) will satisfy this condition.
Alternatively, we may require
\be
\bep = { \varepsilon_\n \over x_\p} > 1
\label{condo}\ee
Recalling that $\varepsilon_\n = (\rho_\p/ \rho_\n )\varepsilon_\p$ and using the relation between the entrainment
and the effective proton mass \eqref{effp}, we
find that \eqref{condo} corresponds to
\be
x_\p > { m_\p^* \over m_\p}
\ee
This condition is unlikely to be satisfied in a neutron star core, where typical values would be
$x_\p \approx 0.1$ and $m_\p^* \approx 0.5 m_\p$, but the possibility is nevertheless interesting.
In particular since the mode will actually be unstable (due to the presence of
the mutual friction) if the condition is met.
Since $\bB>0$ it is clear from \eqref{rcounter} that  the imaginary part of the mode frequency is negative
if \eqref{condo} is satisfied. The existence of this
unstable regime is interesting, at least conceptually.

As a final check let us compare the damping timescale calculated in (\ref{rcounter}) with that calculated using the integral approach, cf. (\ref{tauest}).
Using the definitions for the energy, (\ref{kinetic}), and the dissipation due to mutual friction, (\ref{dE}), one readily finds for the counter-moving r-mode solution; 
\be
\partial_t E_\mathcal{B}=-2\rho(1-x_\p)\mathcal{B}\Omega\int_0^R [l(l+1)-m^2] | u^0_l |^2  r^4 dr
\ee
Meanwhile
\be
E=\frac{1}{2}\rho x_\p(1-x_\p)(1-\bar{\varepsilon})\int_0^R l(l+1) | u^0_l |^2 r^4 dr
\ee
These lead to
\be
\mathrm{Im}\ \omega = \frac{1}{\tau}= \frac{2\Omega\bar{\mathcal{B}}}{(1-\bar{\varepsilon})}\frac{[(l(l+1)-m^2]}{l(l+1)}
\ee
which agrees perfectly with the damping timescale extracted from (\ref{rcounter}).
Since we are using the full dissipative mode-solution in the energy integrals, this is as expected.

\section{Concluding remarks}

The aim of this paper was to lay the foundation for a renewed assault on the problem of
dissipative superfluid neutron star oscillations. We have discussed the oscillations of slowly rotating
superfluid stars, taking into account the mutual friction force at linear order in the (presumed) slow rotation of the star.
We have considered both the
fundamental f-modes and the inertial r-modes. Our analysis differs from previous studies in that we do not assume weak mutual friction from the outset,
the final results are also valid in the strong drag regime.

In the case of the  f-modes, we worked out an analytic approximation for the
mode which allowed us to write down a closed expression for the mutual friction damping timescale.
This result, which is in good agreement with previous numerical results of Lindblom and Mendell \cite{lm95},
 provides useful insight into the dependence on, and relevance of, various equation
of state parameters. The scaling with the harmonic index $l$ is also obvious from our final formula. The
analysis is readily extended to stars with superfluid cores that do not extend all the way to the surface
(as assumed in our analysis), although the result is then less transparent.

In the case of the  r-modes, we have confirmed the existence of two classes of
modes. However, we demonstrated that only one of these sets will remain purely axial in more realistic situations.
This agrees with previous results of Lee and Yoshida \cite{yl1}.
We discussed some peculiarities of the counter-moving r-modes. In particular, the fact that they may be unstable for
some parameter values. Even though we do not expect this instability to be relevant for
realistic superfluid stars, its existence is of conceptual interest.

Building on the formalism and the results presented in this paper, we are currently carrying out a detailed study of the mutual friction damping of the
r-modes at second order in the slow-rotation approximation \cite{bryn}. At the same time we are considering neutron stars with exotic
hyperon and/or quark cores. Since the multifluid aspects of those problems have never been considered in detail, these are exciting developments. 
They are, in fact, necessary if we want to understand the dynamics of realistic models of mature neutron stars.

\acknowledgements
This work was supported by STFC in the UK through grant number PP/E001025/1. KG is supported by the German Science Foundation (DFG) via SFB/TR7.

\end{document}